\documentclass[pdftex]{article}
\usepackage{icrctc07}

\title{SiPM development for Astroparticle Physics applications}
\shorttitle{SiPM development}
\authors{M. Teshima$^{1}$, B. Dolgoshein$^{2}$, R. Mirzoyan$^{1}$, J. Nincovic$^{1}$, E. Popova$^{2}$.}
\shortauthors{M. Teshima and et al.}
\afiliations{$^1$Max-Planck-Insitute for Physics, Foehringer Ring 6, 80805 Munich, Germany\\
$^2$Moscow Engineering Physics Institute, Kashirskoe Shosse 31, 115409 Moscow, Russia}
\email{mteshima@mppmu.mpg.de}

\abstract{
The SiPM is a novel solid state photodetector which can be operated in the single photon counting mode. It has excellent features, such as high quantum efficiency, good charge resolution, fast response, very compact size, high gain of $10^6$, very low power consumption, immunity to the magnetic field and low bias voltage (30-70V). Drawbacks of this device currently are a  large dark current, crosstalk between micropixels and relatively low sensitivity to UV and blue light. In the last few years, we have developed large size SiPMs ($9$ mm$^2$ and $25$ mm$^2$) for  applications in the imaging atmospheric Cherenkov telescopes, MAGIC and CTA, and in the space-borne fluorescence telescope EUSO. The current status of the SiPM development by MPI and MEPhI will be presented.}

\begin{document}
\maketitle
%Begin the section.

\section{Introduction} 

In Astroparticle Physics experiments, we measure the
rare and low flux events, and naturally the instrument will become large in
area and volume. Photodetection techniques play an important role in the
detection of these rare and low flux events. We require low level light
(LLL) detectors working in single photon counting mode. We record the time
profile of signals or require the very fast timing of signals. The imaging of
the signal plays a very important role in Imaging Air Cherenkov Telescopes
(IACTs) for high energy gamma rays and in the air fluorescence technique for Ultra
High Energy Cosmic Rays (UHECRs). 

The SiPM is a solid state photodetector developed for applications in  High Energy Physics, Astroparticle Physics and Medical Science. For Astroparticle Physics applications, the most attractive feature of the SiPM is its high photodetection efficiency ($\sim 70\%$) and the capability of single photon counting. Further, there are many other advantages, such as the compactness of its size and volume, the low bias voltages, the very high gain of $10^6$, the low power consumption, etc.. At present, the major problems of SiPMs are the small size, the cross talk between micropixels in the SiPM and the high dark current.

\begin{figure}
\begin{center}
\includegraphics [width=0.40\textwidth]{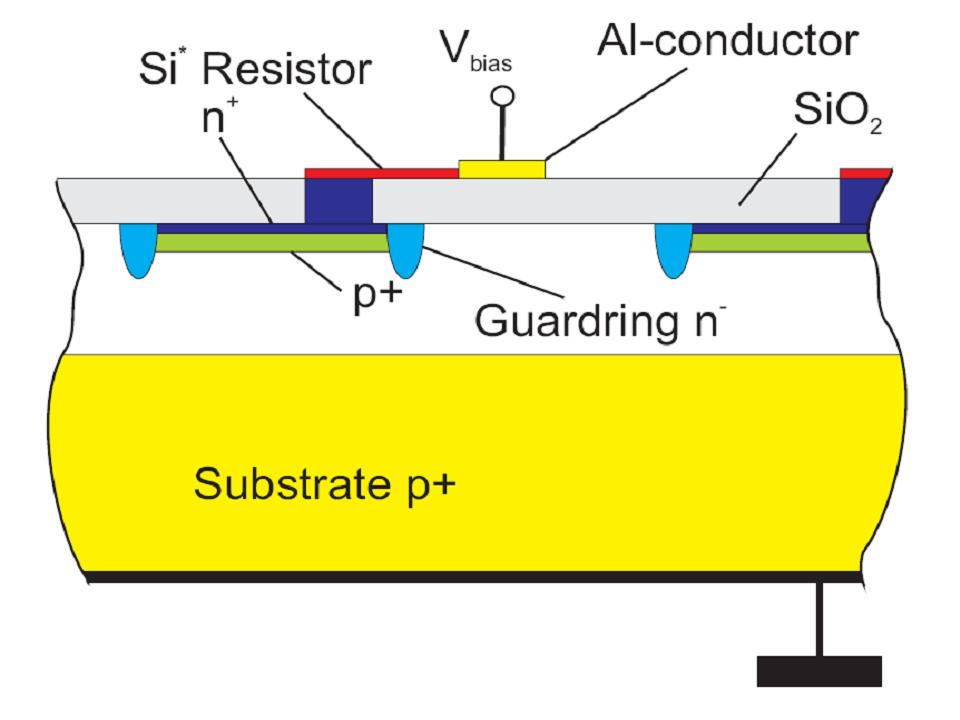}
\vspace{-20pt}
\end{center}
\caption{Structure of the micropixel of a SiPM.}
\label{fig1}
\end{figure}

For the MAGIC telescope [1], the high photodetection efficiency will allow us to lower the threshold energy of gamma ray detection down to 10-20 GeV. We can also design new smaller and inexpensive Cherenkov telescopes keeping the performance at an equally high level. This allows us to consider an array of smaller high performance telescopes at a low budget. For MAGIC, the pixel size of the photodetector shall be a $10-20$ mm size (corresponding to $0.05 - 0.1^\circ$ with a winston cone). Even if we consider smaller pixels below $0.05^\circ$, we do not gain in angular resolution and gamma/hadron separation. 

\begin{figure}
\begin{center}
\includegraphics [width=0.48\textwidth]{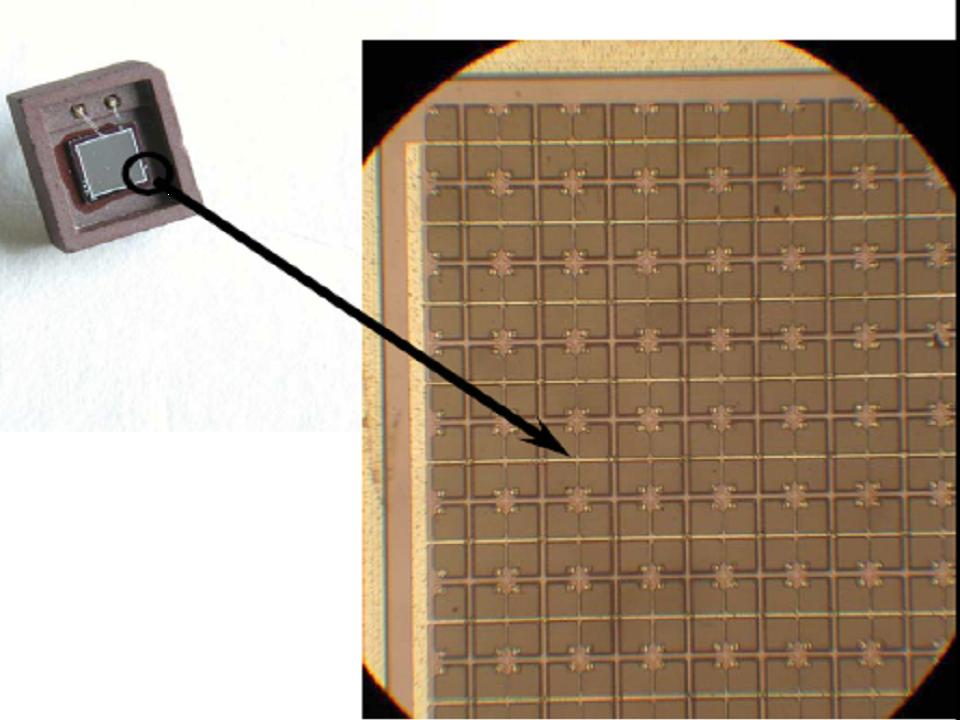}

\vspace{-8pt}
\end{center}
\caption{Microscope picture of a $ 9(3 \times 3)$ mm$^2$ SiPM. It consists of an array of $5625(75 \times 75)$ micropixels with $30 \times 30 $ $\mu$m$^2$ size.}
\label{fig2}
\end{figure}

For the space-borne air fluorescence detector EUSO[3], photodetectors with a size of $3-5$ mm are required for $0.1^\circ$ pixel imaging. The high photodetection efficiency will ensure the detection of UHECRs above $(2-3) \times 10^{19}$ eV and allow a large overlap in the energy range with ground based detectors, Auger[4] and TA[5]. 
Alternatively it will allow us to use the high altitude orbit or tilted mode to cover larger detection volumes for UHECRs. Characteristics like the low power consumption and insensitivity to magnetic fields are important advantages for space detectors.

Therefore, the larger size SiPM of $25$ mm$^2$ or even $100$ mm$^2$, with high photondetecion efficiency, low dark rate, low crosstalk and enhanced sensitivity in UV and blue light is the prime target of our SiPM development. 
Silicon itself has the nice feature of nearly $100\%$ internal quantum efficiency. The increase of the filling factor of the sensitive area on the device surface and the increase of the Geiger efficiency will give us a very high photodetection efficiency.

\begin{figure}
\begin{center}
\includegraphics [width=0.48\textwidth]{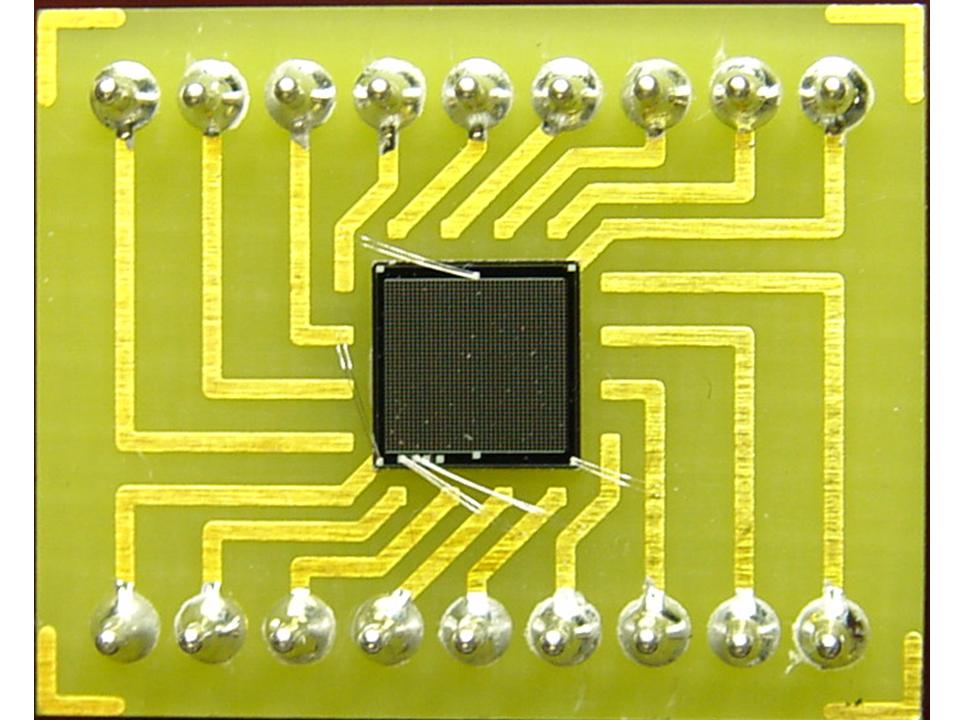}
\vspace{-20pt}
\end{center}
\caption{$25 (5 \times 5)$ mm$^2$ SiPM. It consists of an array of $1600(40 \times 40)$ micropixels with $100 \times 100$ $\mu$m$^2$ size.}
\label{fig3}
\end{figure}

\section{Structure and operation of the SiPM}

The SiPM consists of an array of a large number of micropixels (avalanche
diodes) working in Geiger mode. The schematic picture of the individual
micropixel's structure is shown in figure \ref{fig1}. This micropixel works
somehow in digital mode, giving us the signal $Q = C_{pixel} \times (V -
V_{bd})$ in each Geiger discharge. Here, $C_{pixel}$ is the capacitance between
anode and cathode of the avalanche diode, $V_{bd}$ is the breakdown voltage of
micropixel (avalanche diode), $V - V_{bd}$ is called over voltage. The typical
$Q$ is $10^6 e$. The anodes of all micropixels are joined together on a common
substrate and are working on common load in case of n on p structure.

The operation of the individual micropixel is as follows. The micropixel is
biased with a higher voltage than the breakdown voltage. This over voltage is
typically $10\%$ or $20\%$ of the breakdown voltage. The micropixel is then kept
in a meta-stable condition, and if a pair of electron and hole is created in
the depletion or drift region by an incident photon, a Geiger discharge will
be triggered. The current flow from the biasing circuit through the quenching
register causes the bias voltage to drop and the Geiger process will stop
automatically. The biasing circuit will recharge the capacitance of the SiPM to
raise the bias voltage again up to the meta-stable over voltage. This recharge
takes typically $1$ $\mu$sec. The dark rate (night sky back ground, or dark
current) DR per micropixel of $30 \times 30$ $\mu$m$^2$ or $100 \times 100$ $\mu$m$^2$ is usually low enough, i.e. $1-10$ kHz, even at room temperature, so the dead time
and the decrease of photodetection efficiency can be negligibly small.

\section{Development of SiPMs for MAGIC, CTA and EUSO}

MPI and MEPhI have developed SiPMs with a size of $ 9 (3 \times 3)$ mm$^2$ and $ 25 (5 \times 5)$ mm$^2$ for MAGIC [1], CTA [2] and EUSO [3] applications as shown in figure \ref{fig2} and figure \ref{fig3}[6,7,8,9].
The micropixel size is enlarged to $100 \times 100$ $\mu$m$^2$ to get a higher filling factor of the sensitive region, in case of the $25$ mm$^2$ SiPM. 
The devices were manufactured by the company Pulsar in Moscow.

\begin{figure}[h]
\begin{center}
\
\includegraphics [width=0.48\textwidth]{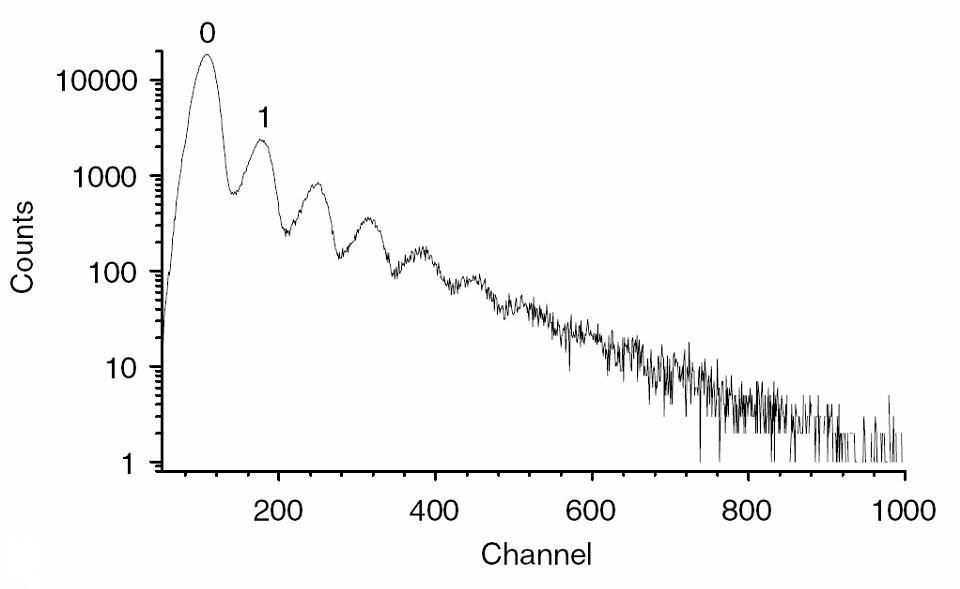}

\includegraphics [width=0.48\textwidth]{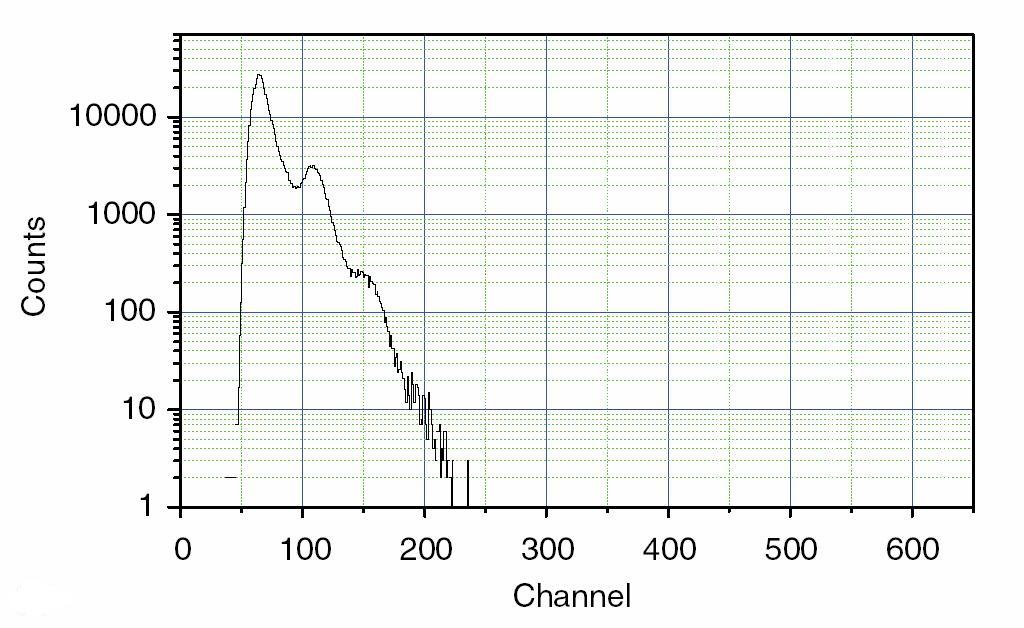}
\vspace{-24pt}

\end{center}
\caption{Top and bottom panels show the pulse height distributions of dark current without and with OC suppression structure, respectively. They are operated with the gain of $3 \times 10^6$ and $3 \times 10^7$. In both panels, the first peak is pedestal, the second and third one are single and two ph.e., respectively.}
\label{fig4}
\end{figure}

\subsection{Dark Rate}
The dark rate of SiPMs is about $ 1-2$ MHz/mm$^2$ at room temperature ($\sim 20^\circ C$). Dark current is produced by two mechanisms, i.e. the electric field assisted tunneling effect and the thermal effect. The dark current can be reduced by one or two orders of magnitude by cooling the temperature down to $- 7^\circ C$ or $-35^\circ C$ in our devices, respectively. 
In our device the main component of dark current is the thermal one. 
The clever design  avoids unnecessary extremely high electric fields will reduce the electric field assisted tunneling. The usage of ultra pure silicon wafers, the reduction of the volume of depletion and drift regions will help to reduce the dark rate.

\subsection{Optical Crosstalk} The Optical Crosstalk (OC) is a specific
phenomenon in SiPMs. The Geiger avalanche in one micropixel will provoke the
emission of light, typically $\sim 10$ photons/avalanche of $10^6 e$, and these
photons will hit other pixels. This means that the single ph.e. has a chance to
produce a multiple ph.e. signal.  The emission of light is proportional to the
Geiger gain. The Geiger efficiency which determines the final photodetection
efficiency has also a strong positive dependence on the Geiger gain. Therefore,
if we want to obtain a high photodetection efficiency, we need to operate SiPMs
in high Geiger gain. This condition will introduce severe optical crosstalk in
SiPMs. To avoid this problem we have introduced the OC suppression structure.

Optical crosstalk entails a serious problem in our specific usage, because we use photodetectors not only to record photon intensity but also to trigger events. Especially in the Cherenkov experiment and air fluorescence experiment, we have a high rate of background photons from the night sky. Even if there are a series of single ph.e., this optical crosstalk produces multiple ph.e. signals with a high rate and we are consequently required to effectively set the trigger threshold high in order to obtain a reasonable trigger rate. 

The top panel of figure 4 shows the pulse height distribution of dark rate in the standard SiPM without the OC suppression structure. We can see the exponential long tail in the multiple ph.e. region. These large pulses will raise the trigger threshold like after pulses in PMTs.  By manufacturing the OC suppression structure (the groove structures between micropixels which block the transmission of light between micropixels through silicon), we succeed to suppress these unfavorable optical crosstalk. As shown in the bottom of figure 4, the optical crosstalk level was significantly reduced by this improvement. We can evaluate this improvement with the parameter of the Excess Noise Factor (ENF); the ENF of about 1.6 was improved to ${\rm ENF} = 0.97 \pm 0.05$ with the OC suppression structure.

\begin{figure}
\begin{center}
\
\includegraphics [width=0.48\textwidth]{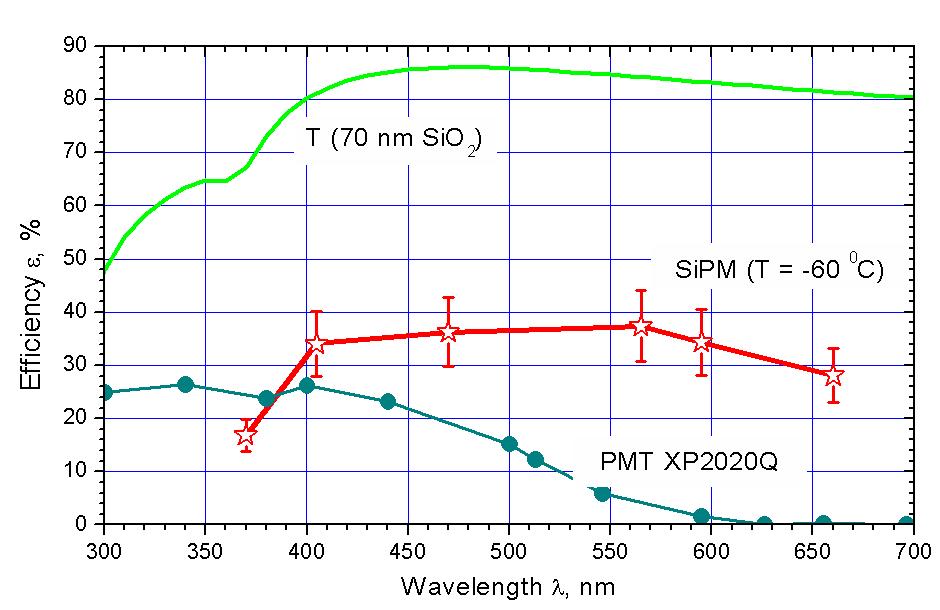}
\
\vspace{-24pt}
\end{center}
\caption{The photodetection efficiency of SiPM of $25(5 \times 5)$ $\mu$m$^2$ with micropixels size of $100 \times 100$ $\mu$m$^2$(filling factor of $64\%$) at a temperature of $-60^\circ C$ is compared with that of a Photomultiplier.}
\label{fig5}
\end{figure}

\begin{figure}
\begin{center}
\
\includegraphics [width=0.48\textwidth]{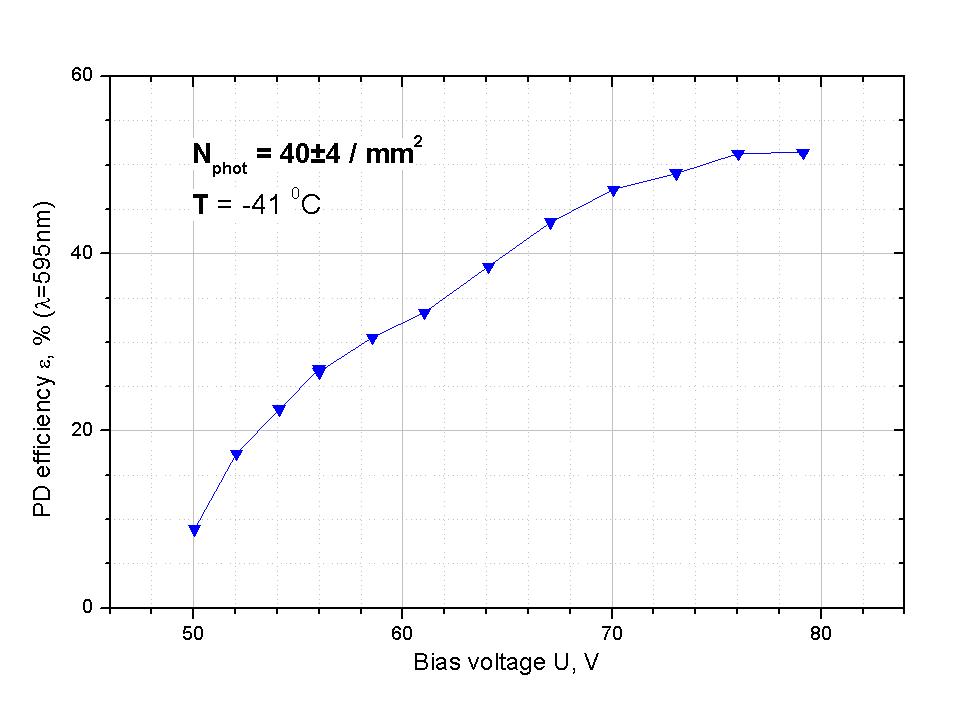}
\vspace{-32pt}
\end{center}
\caption{The photodetection efficiency of SiPM of $25(5 \times 5)$ $\mu$m$^2$ with micropixels size of $100 \times 100$ $\mu$m$^2$(filling factor of 64\%) at large overvoltages.}
\label{fig6}
\end{figure}

\subsection{Photodetection efficiency}
Photodetection efficiency PDE is determined by four factors, i.e. 
the light transmission from surface to depletion or drift layers $T$, 
the filling factor of the sensitive area, (sensitive area)/(all area) $F$, 
the quantum efficiency of silicon $QE(\lambda)$, 
and the Geiger efficiency $G$.

$$ {\rm PDE} = T \times F \times QE(\lambda) \times G .$$

Silicon has a high reflectivity index $n \sim 4.0$, and with AR coating we can increase the light transmission up to a $90-95\%$ level. As shown in figure \ref{fig1}, there are guard rings, quenching registers, aluminum conductors for bias voltage supply around micropixels. This structure will limit the filling factor to $80\%$. The internal quantum efficiency of silicon $QE(\lambda)$ is close to $100\%$, and the Geiger efficiency can be raised up to $\sim 90\%$ in high gain operation (high over voltage).  
Theoretically, the total PDE can be increased up to $\sim 70\%$ with the existing technologies.

The PDE of a $9$ mm$^2$ SiPM is about  $25\%$ between 450nm and 550nm[6]. This $9$ mm$^2$ SiPM consists of $5625 (= 75 \times 75)$ micropixels with a size of $30 \times 30$ $\mu$m$^2$ (filling factor of $\sim 50\%$). 

The SiPM of of $25$ mm$^2$ consists of 1600 (40x40) micropixels with the size of $100 \times 100$ $\mu$m$^2$ (the filling factor of $64\%$). The PDE of such a structure is shown in figure 5. For large overvoltages where the Geiger efficiency is close to $100\%$, the PDE of a $100 \times 100$ $\mu$m$^2$ structure reaches $\sim 50 \%$ as shown in figure 6.

\vspace{8pt}

\noindent{\bf Conclusion}

We have successfully developed $9$ mm$^2$ and $25$ mm$^2$ SiPMs. Their characteristics are almost satisfactory, but we consider further developments/improvements, the enhancement of UV-Blue light sensitivity with reverse structure (p on n), and larger SiPMs of $100$ mm$^2$ size or the array of $4 \times 4$ SiPMs of $25$ mm$^2$ size for MAGIC [1], CTA [2] and EUSO [3] applications. 

\vspace{4pt}

{\small
\noindent{\bf Acknowledgements}
We would like to acknowledge the support by Max-Planck-Society. \\
\noindent {\bf Reference} \\
\noindent [1] http://wwwmagic.mppmu.mpg.de/ \\
\noindent [2] see contributions in this conference \\
\noindent [3] http://euso.riken.go.jp/ \\
\noindent [4] http://www.auger.org/ \\
\noindent [5] http://www-ta.icrr.u-tokyo.ac.jp/ \\
\noindent [6] B. Dolgoshein et al., NIM A563 (2006) 368. \\
\noindent [7] R. Mirzoyan et al., N.I.M. A572 (2007) 493. \\
\noindent [8] P. Buzhan et al., N.I.M. A567 (2006) 78. \\
\noindent [9] A.N. Otte et al., N.I.M. A 545 (2005) 705. \\
}

\end{document}